\documentclass{icrc2009}

\usepackage{graphicx}   
\usepackage[font=footnotesize caption=false]{subfig} 
\usepackage{fixltx2e}
\usepackage{url}

\newcommand{\shorttitle}[1]%
{\markboth{Proceedings of the 31\MakeLowercase{$^{st}$} ICRC, {\L}\'{o}d\'{z} 2009}{#1} }
\newcommand{\etal}{\MakeLowercase{\textit{et al. }}} 

\newcommand{\gray}[1]{$\gamma$-ray{#1}}

\newcommand{\gardian}{{\it GaRDiAn}{}}

\newcommand{\Xco}{$X_{\rm CO}$}

\newcommand{\fermi}{{\it Fermi}{}}

\begin{document}
\title{\fermi{} LAT Measurements of the Diffuse Gamma-Ray Emission at Intermediate Galactic Latitudes}

\author{\IEEEauthorblockN{Troy A. Porter\IEEEauthorrefmark{1} for the {\em Fermi} LAT Collaboration}
                            \\
\IEEEauthorblockA{\IEEEauthorrefmark{1}Santa Cruz Institute for Particle Physics, University of California, 1156 High Street, Santa Cruz 95064}}

\shorttitle{Porter \etal LAT Measurements of Mid-Latitude Diffuse Emission}
\maketitle

\begin{abstract}
The diffuse Galactic \gray{} emission is produced by cosmic rays (CRs)
interacting
with the interstellar gas and radiation field.
Measurements by the Energetic Gamma-Ray Experiment Telescope (EGRET) 
instrument on the {\em Compton Gamma-Ray Observatory (CGRO)} indicated
excess \gray{} emission $\geq 1$ GeV relative to 
diffuse Galactic \gray{} emission models 
consistent with directly measured CR spectra 
(the so-called ``EGRET GeV excess''). 
The excess emission was observed in all directions on the sky,
and a variety of explanations have been proposed,
including beyond-the-Standard-Model scenarios like annihilating or decaying 
dark matter.
The Large Area Telescope (LAT) instrument on the \fermi{} 
Gamma-ray Space 
Telescope 
has measured the diffuse 
\gray{} 
emission with unprecedented sensitivity and resolution.
We report on LAT measurements of the diffuse \gray{} 
emission for energies 100 MeV to 10 GeV and Galactic latitudes
$10^\circ \leq |b| \leq 20^\circ$.
The LAT spectrum for this region of the sky is well reproduced by the 
diffuse Galactic \gray{} emission models 
mentioned above and inconsistent with the EGRET GeV excess.
\end{abstract}

 
\section{Introduction}

The diffuse \gray{} emission, both Galactic and extragalactic, is of
significant interest for astrophysics, particle physics, and cosmology.
The diffuse Galactic emission (DGE) is produced by interactions of 
CRs, mainly protons and electrons, 
with the interstellar gas (via $\pi^0$-production and Bremsstrahlung) and 
radiation field (via inverse Compton [IC] scattering).
It is a direct probe of CR fluxes in distant locations, and
may contain signatures of physics beyond the Standard Model, 
such as dark matter annihilation or decay.
The DGE is a foreground for point-source detection and hence influences the 
determination of their positions and fluxes. 
It is also a foreground for the much fainter extragalactic
component, which is the sum of contributions from unresolved sources
and truly diffuse
emission, including any signatures of large scale structure formation,
emission produced by 
ultra-high-energy CRs interacting with
relic photons, and many other processes (e.g.,~\cite{Dermer2007} and references
therein).
Therefore, understanding the DGE is a necessary first step in all such studies.

The 
excess diffuse emission $\geq 1$ GeV
in the EGRET data \cite{Hunter1997} relative to that expected from
DGE models consistent with the directly measured CR nucleon and 
electron spectra \cite{Strong2000} led to the 
proposal that this emission was the 
long-awaited signature of dark matter annihilation~\cite{deBoer2005}.
More conventional interpretations included variations of CR
spectra in the Galaxy~\cite{Porter1997,Strong2000}, 
contributions by unresolved point sources
\cite{BV2004}, and instrumental 
effects~\cite{Hunter1997,Moskalenko2007}.

A model of the DGE depends on the CR spectra throughout the 
Galaxy as well as the distribution of the target gas and interstellar 
radiation field (ISRF).
Starting from the distribution of CR sources and particle injection spectra,
the distribution of CRs throughout the Galaxy is determined taking into 
account relevant energy losses and
gains, then 
the CR distributions are folded with the target
distributions to calculate the 
DGE \cite{SMR04}.  
Defining the inputs and calculating the models are not trivial
tasks and involve analysis of data from
a broad range of astronomical and astroparticle
instruments~\cite{StrongAnnRev2007}.

The \fermi{} LAT was launched on June 11, 2008.
It is over an order of magnitude more sensitive than its 
predecessor, EGRET, with a more stable response due to the lack of 
consumables.
The LAT data permit more detailed studies of the DGE than have been 
possible ever before. 

In this paper, analysis and results for the DGE are shown for
the Galactic mid-latitude
range $10^\circ \leq |b| \leq 20^\circ$ measured
by the LAT in the first 5 months of the science phase of the mission.
This region was chosen for initial study since it maximises the fraction
of signal from DGE produced within several kpc of the Sun and 
hence uncertainties associated with CR propagation, knowledge
of the gas distribution, etc., should be minimised.
At lower Galactic latitudes the large-scale DGE is dominant 
while the
emission at higher latitudes is more affected by contamination from 
charged particles misclassified as photons and uncertainties in the model 
used to estimate the DGE.
Further details of the present analysis are given in \cite{GeVExcesspaper}

\section{LAT Data Selection and Analysis}

The LAT is a pair-conversion telescope with a precision tracker and 
calorimeter, each consisting of a $4\times 4$ array of 16 modules, a segmented 
anti-coincidence detector (ACD) that covers the tracker array, 
and a programmable 
trigger and data acquisition system. 
Full details of the instrument, onboard and ground data processing, and other 
mission-oriented support are given in \cite{InstrumentPaper}.

The 
data selection used in this paper is made using the standard 
LAT ground processing and background rejection scheme.
This consists of two basic parts: first a simple accept-or-reject selection 
(prefiltering) followed by a classification tree (CT) \cite{1984CTbook} 
based determination of the relative probability of being background or 
signal. 
The prefiltering phase screens particles entering the LAT
for their charge neutrality using the tracker and ACD.
The direction reconstruction software extrapolates found particle trajectories
in the tracker back to the scintillation tiles of the ACD covering the LAT 
and only accepts events in which the intersected tile shows no significant 
signal.
In addition, considerations such as how well the found tracks project into 
the measured energy centroid in the calorimeter, and the shape of the shower 
energy deposition, are also used.   
The overall background rejection of the prefiltering phase is $10^3 - 10^4$
depending on energy, 
yielding a \gray{} efficiency $> 90\%$ relative to \gray{s} that convert in
the LAT.

Classification trees, which afford an 
efficient and 
statistically robust method for distinguishing signal from noise, are used to 
reduce backgrounds further.
The CTs are trained using Monte Carlo data which have passed the prefiltering 
described above.
Multiple CTs are used to make the procedure robust 
against statistical fluctuations during the training procedure. 
The result from averaging these CTs is the probability for an event to
be a photon or background.
This final rejection parameter allows the background levels to be 
set according to the needs of the analysis.
For the analysis of diffuse emission, the CT generated probability is set 
to allow a Monte Carlo predicted orbit-averaged 
background rate of $\sim0.1$ Hz integrated over
the full instrument acceptance $> 100$ MeV.
This yields a \gray{} efficiency $> 80\%$, and the remaining background 
is at a level where the majority of the contamination arises from irreducible
sources such as 
\gray{s} produced 
by CR interactions in the passive material outside the ACD, e.g., 
the thermal blanket and micrometeroid shield for the LAT.
The events corresponding to the above criterion are termed ``Diffuse'' class and
are the standard low-background event selection.

The analysis presented here uses post-launch instrument response functions
(IRFs).
These take into account pile-up and accidental coincidence 
effects in the detector 
subsystems that are not considered in the definition of the pre-launch IRFs. 
Cosmic rays, primarily protons,
pass through the LAT at a high rate and sufficiently 
near coincidences with \gray{s}
leave residual signals that can result in \gray{s} being misclassified,
particularly at energies $\leq 300$ MeV.  
The post-launch IRFs were
derived using LAT events read out at regular intervals 
as a background overlay on
the standard simulations of \gray{s} and provide an accurate
accounting for the instrumental pile-up and accidental coincidence effects.  
The on-axis effective area for the event selection used in this paper is 
$\sim 7000$ cm$^2$ at 1 GeV and is energy dependent; this is approximately 10\%
lower than the pre-launch effective area corresponding to the 
same event selection.
The systematic uncertainties of the effective area, evaluated
by comparing the efficiencies of analysis cuts for data and simulation of
observations of Vela, are also energy dependent:  10\% below 100 MeV, 
decreasing to
5\% at 560 MeV, and increasing to 20\% at 10 GeV and above.  
The point spread function (PSF) and energy resolution are as described in 
\cite{InstrumentPaper}.

The LAT nominally operates in a scanning mode that covers the whole sky 
every two orbits (i.e., ~3 hrs). 
We use data taken in this 
mode from the commencement of scientific operations in 
mid-August 2008 to the end of December 2008. 
The data were prepared using the LAT Science Tools 
package, 
which is available from the \fermi{} Science Support 
Centre.
Events satisfying the Diffuse class selection and coming from 
zenith angles $< 105^\circ$ (to greatly reduce the contribution by 
Earth albedo
\gray{s}) were used. 
To further reduce the effect of Earth albedo 
backgrounds, the time intervals when the Earth was appreciably within the 
field of view (specifically, when the centre of the field of view was 
more than 47$^\circ$ from the zenith) were excluded from this analysis.
This leaves 9.83 Ms of total livetime in the data set.
The energy-dependent exposure was calculated using the IRFs described above.

The photon counts and exposure were further processed using the 
\gardian\ package, part of a suite of tools we have developed to 
analyse the DGE~\cite{gadget_and_gardian}.
Gamma-ray skymaps with 5 bins per decade in energy from 100 MeV to 10 GeV 
were generated.
For each energy bin the intensity was obtained by dividing the in-bin counts by 
the spectrally-weighted exposure over the bin.
We used two methods for the spectral weighting: a power law with index $-2$ and
the spectral shape of the assumed DGE model (described below).
With the energy binning used in this paper the differences in the derived 
intensities were $< 1$\% between these two weighting schemes.


\begin{figure*}[!t]
  \centerline{
    {\includegraphics[width=7.5cm]{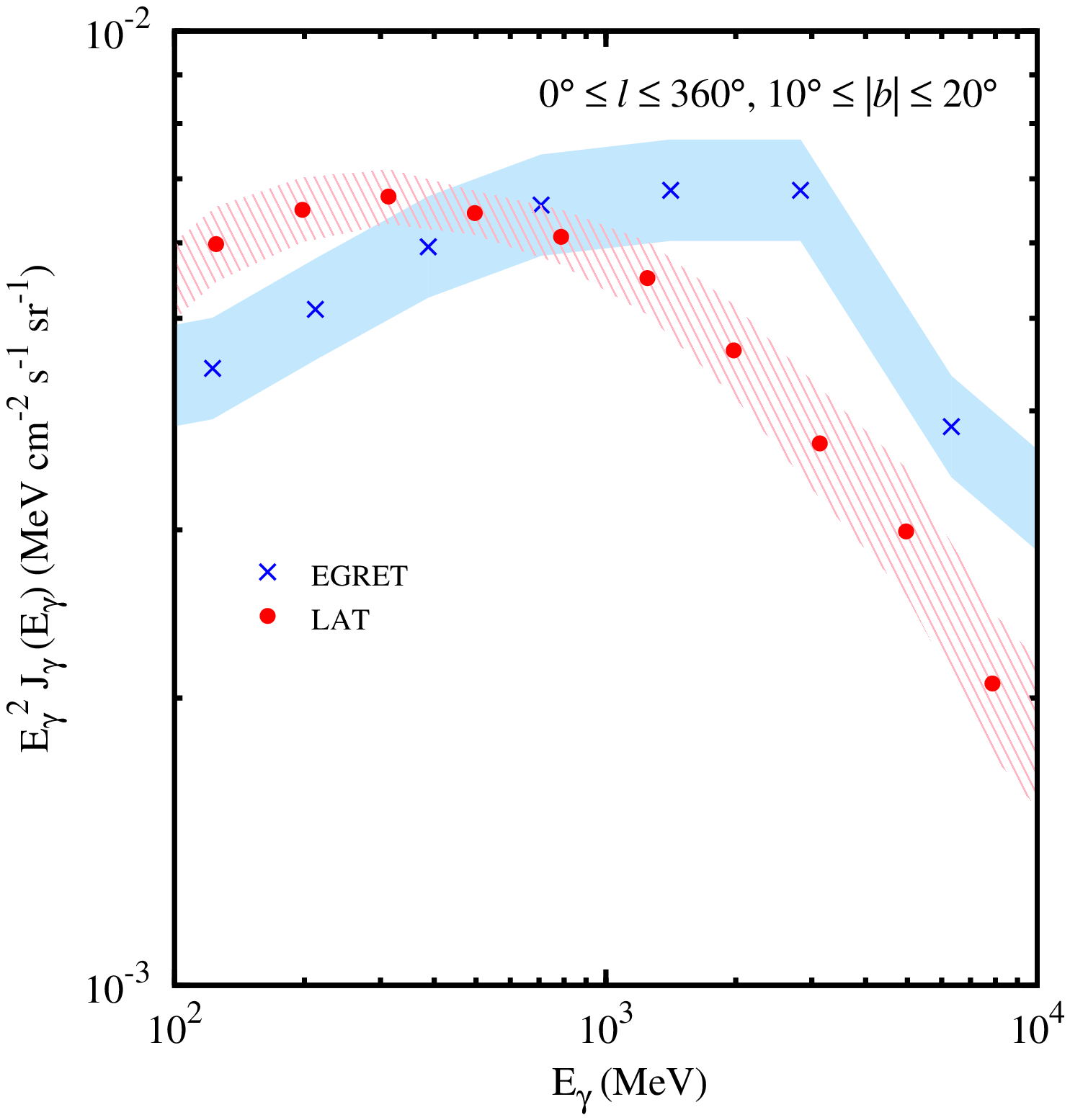}}
    {\includegraphics[width=7.5cm]{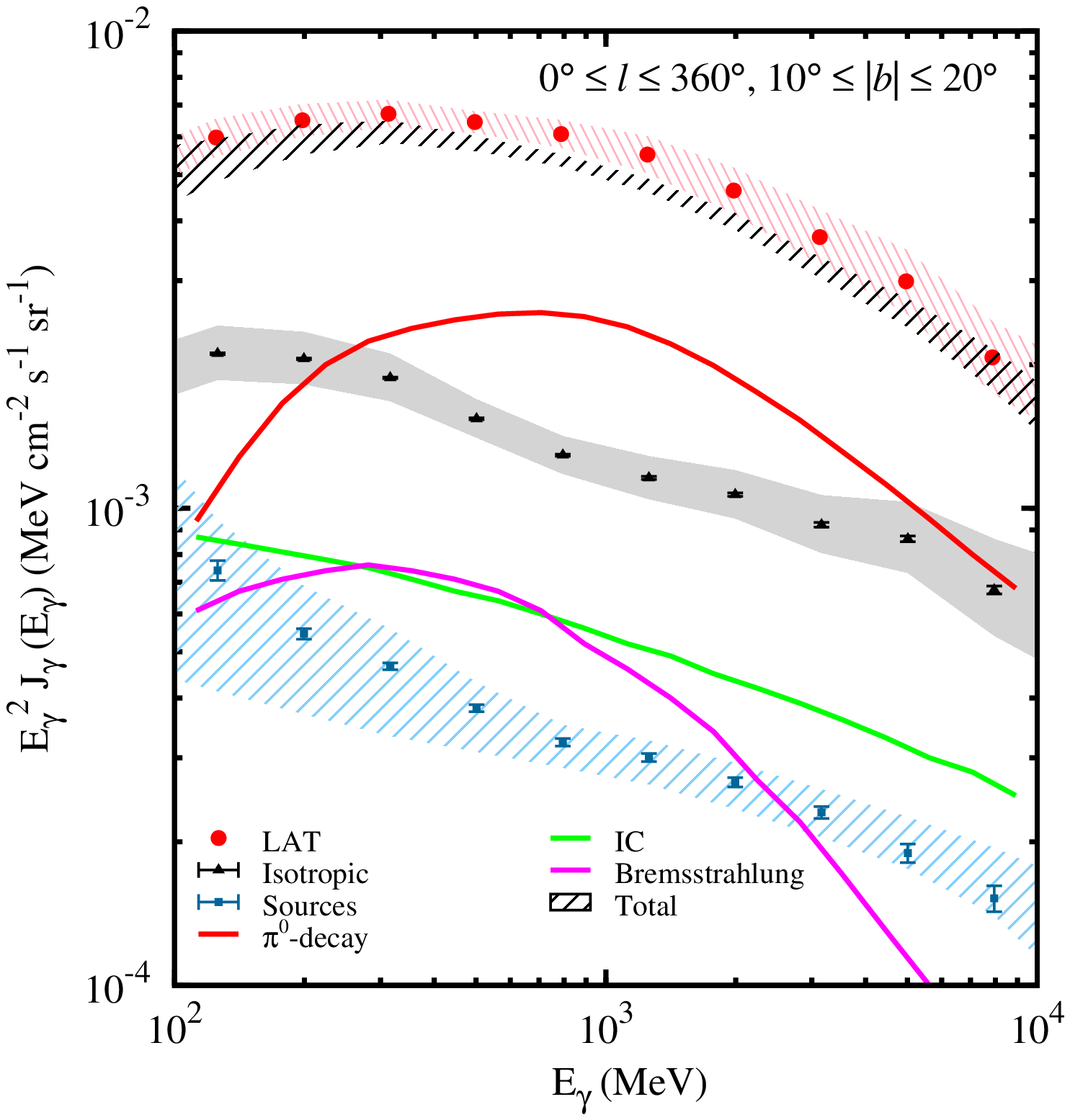}}
  }
  \caption{{\it Left:} Preliminary diffuse emission intensity averaged over all Galactic
longitudes for latitude range $10^\circ \leq |b| \leq 20^\circ$. 
Data points: LAT, red dots; EGRET, blue crosses.
Systematic uncertainties: LAT, red; EGRET, blue.
{\it Right:} Preliminary LAT data with model, source, and 
UIB components for same sky region.
Model (lines): $\pi^0$-decay, red; Bremsstrahlung, magenta; IC, green.
Shaded/hatched regions: isotropic, grey/solid; source, blue/hatched; 
total (model + UIB + source), black/hatched.}
  \label{fig1}
\end{figure*}


Figure~\ref{fig1} (left) shows 
the LAT data averaged over 
all Galactic longitudes
and the latitude range 
$10^\circ \leq |b| \leq 20^\circ$.
The hatched band surrounding the LAT data indicates the systematic 
uncertainty in the measurement due to the uncertainty in the effective area 
described above.
Also shown are 
the EGRET data for the same region 
of sky
derived
from count maps and exposures 
available via the {\em CGRO} Science Support Centre
\footnote{http://heasarc.gsfc.nasa.gov/docs/cgro/egret/} and processed 
following the procedure described in \cite{SMR04} and we have included
the standard systematic uncertainty of 13\% \cite{Esposito1999}.
For both data sets the contribution by point sources has not been subtracted.
The LAT-measured spectrum is significantly softer than the EGRET measurement 
with an integrated intensity $J_{\rm LAT}(\geq 1 \, {\rm GeV}) = 
2.35\pm0.01\times10^{-6}$ cm$^{-2}$ s$^{-1}$ sr$^{-1}$ 
compared to the EGRET integrated intensity 
$J_{\rm EGRET}(\geq 1\, {\rm GeV}) = 3.16\pm0.05\times10^{-6}$ 
cm$^{-2}$ s$^{-1}$ sr$^{-1}$ where the errors are statistical only.
Not included in the figure is the systematic uncertainty in the energy scale, 
which is conservatively estimated from comparison between Monte Carlo and 
beam test data as $< 5$\% for 100 MeV to 1 GeV, and $< 7$\% above 1 GeV 
where it is believed that if any bias is present energies are overestimated.
Taking the uncertainty on the energy scale into account, the LAT spectrum could
be softer, increasing the discrepancy with the EGRET spectrum further.

Figure~\ref{fig1} (right) compares the LAT spectrum with the 
spectra of an {\em a priori} DGE model, and a point-source 
contribution and unidentified background (UIB) component 
derived from fitting the LAT data that are described
below.
The DGE model is an updated version of the ``conventional'' 
model from GALPROP \cite{SMR04}. 
Major
improvements include use of the formalism and corresponding code for pion 
production in $pp$-interactions by
\cite{Kamae2006}, 
a complete recalculation of the ISRF \cite{Porter2008}, updated gas maps,
and an improved line-of-sight integration routine.

The source and UIB components were obtained by fitting 
the LAT data using \gardian\ with the DGE model fixed.
Point source locations were taken from the 3 month \fermi{} LAT source list
down to 
sources with 5-$\sigma$ significance.
Due to the limited statistics of all but the very brightest sources, we used 
3 bins per energy decade in the fitting procedure. 
Source positions were fixed but the spectra were fit using one free 
parameter per energy bin.
The UIB component was determined by fitting the data
and sources over all Galactic longitudes for the high-latitude region 
$|b| \ge 30^\circ$ for the full LAT energy range shown in the figure.
Using this high-latitude region minimises the effect of contamination by the 
bright Galactic ridge which can be significant even up to $\sim 10^\circ$ 
from the plane due to the long tails of the PSF at low energies.

To determine the uncertainty of the
source and UIB components, we modified the effective 
area to the
extremes of its systematic uncertainty defined before and refitted the data.
Since the DGE model is fixed, 
the absolute change in intensity caused by the modification to the effective
area propagates directly to the source and UIB components.
The systematic uncertainty on these components is energy dependent and due to
several effects.

For energies $\geq 10$ GeV the PSF is $\sim 0.2^\circ$ (68\% containment) and
the sources are well-localised spatially.
Since the model is fixed and the sky maps are sparser at high latitudes for the 
data taking period in this paper, the UIB component 
absorbs almost
all of the intensity from the modification to the 
effective area.
At low energies the PSF is wider, $3.5^\circ$ (68\% containment) at 100 MeV for
\gray{} conversions in the front section of the LAT, and the 
sources are less well-localised spatially.
In addition, the sky maps are well populated even at high latitudes 
and display spatial structure.
The PSF broadening of the sources provides spatial structure and 
because the DGE model is fixed, more intensity is assigned to the source 
component to compensate in the fit.
These effects lead to the systematic error in the source component
being relatively larger than the isotropic at low energies and vice
versa at high energies.
Note, this applies for the high-latitude region from where the UIB 
component is derived, and also for the mid-latitude range for which we show
the combined contribution by sources in Fig.~\ref{fig1} (right).
Because the uncertainties in the source and UIB components 
are not independent we have conservatively added their systematic uncertainties
for the total intensity band shown in Fig.~\ref{fig1} (right).

The UIB component comprises the
true extragalactic diffuse \gray{} emission, 
emission from unresolved Galactic and extragalactic sources, and 
residual particle 
backgrounds (CRs that pass the \gray{} classification analysis 
and \gray{s} produced by CR
interactions in the passive material outside the ACD) in the LAT data.
In addition, other relevant 
foreground components that are not completely 
modelled, such as emission from the 
solar disk and extended emission \cite{Moskalenko2006,Orlando2007,Orlando2008} and
other potentially relevant 
``diffuse'' sources \cite{moonpaper,sssbpaper,oortpaper} are included.
Hence, the UIB component does {\em not} 
constitute a measurement of the extragalactic diffuse emission. 
Furthermore, comparison with the EGRET estimate of the extragalactic diffuse
emission \cite{sreekumar1998} is problematic due to the different DGE models
used and analysis details that are beyond the scope of the current paper and
will be addressed in a subsequent publication.

\section{Discussion}
The intensity scales of the LAT and EGRET have been found to be different with
the result that the LAT-measured spectra are softer.
In our early study of the Vela spectrum \cite{velapaper}, which was made 
using pre-launch IRFs, the difference was apparent already above 1 GeV.
Following on-orbit studies new IRFs have been developed to account for 
inefficiencies in the detection of \gray{s} in the LAT due to pile-up and 
accidental coincidence effects in the detector subsystems.
The inefficiency increases at lower energies, with the result that the IRFs 
used in the present analysis indicate greater intensities in the range below 
1 GeV, with the magnitude of the effect ranging up to 30\% at 100 MeV.
Our confidence that the IRFs used in the present analysis accurately represent
our knowledge of the instrument comes from detailed instrument simulations that 
were validated with beam tests of calibration units, and to post-launch 
refinements based on actual particle backgrounds.
The systematic uncertainty on the effective area gives an energy dependent
measure of our confidence in the IRFs used in the present analysis.

As a consequence, 
the LAT-measured DGE spectrum averaged over all Galactic longitudes for the 
latitude range $10^\circ \leq |b| \leq 20^\circ$ is systematically softer 
than the EGRET-measured spectrum.
The spectral shape is compatible with that of an {\em a priori} DGE model that 
is consistent with directly measured CR spectra.
The excess emission above 1 GeV measured by EGRET is not seen by the LAT 
in this region of the sky.

While the LAT spectral shape is consistent with the DGE model used in this 
paper, the overall model emission is too low thus giving rise to a 
$\sim10-15$\% excess over the energy range 100 MeV to 10 GeV.
However, the DGE model is based on pre-\fermi{} data and knowledge
of the DGE.
The difference between the model and data is of the same order as the 
uncertainty in the measured CR nuclei spectra at the relevant 
energies \cite{besspaper}.  
In addition, other model parameters that can affect the \gray{} production 
rate (e.g., \Xco –- the conversion between CO line intensity and molecular
hydrogen column density in the interstellar medium) have not been modified 
in the present paper.
Overall, 
the agreement between the LAT-measured spectrum and the model shows that 
the fundamental processes are consistent with our data, thus providing a
solid basis for future work understanding the DGE.

{\it Acknowledgements:}
The \fermi{} LAT Collaboration acknowledges support from a number of 
agencies and institutes for both development and the operation of the LAT 
as well as scientific data analysis. 
These include NASA and DOE in the United States, CEA/Irfu and IN2P3/CNRS 
in France, ASI and INFN in Italy, MEXT, KEK, and JAXA in Japan, and 
the K.~A.~Wallenberg Foundation, the Swedish Research Council and the 
National Space Board in Sweden. 
Additional support from INAF in Italy for science analysis during the 
operations phase is also gratefully acknowledged.

GALPROP development is funded via NASA grant NNX09AC15G.

\end{document}